\documentclass[preprint,showpacs,preprintnumbers,amsmath,amssymb]{revtex4}
\usepackage{graphicx}
\usepackage{dcolumn}
\usepackage{bm}

\begin{document}
\vphantom{math}
\vskip1.1cm

\title{ON EVOLUTION EQUATIONS OF QUANTUM-CLASSICAL SYSTEMS}%

\author{V.I. Gerasimenko}%
\email{gerasym@imath.kiev.ua}
\affiliation{Institute of Mathematics NASU,\\3, Tereshchenkivs'ka str., \\01601 Kyiv, Ukraine}

\begin{abstract}
We consider the links between consistent and approximate descriptions of the
quantum-classical systems, i.e. systems are composed of two interacting subsystems, one
of which behaves almost classically while the other requires a quantum description.
\end{abstract}



\maketitle

\newpage
\vphantom{math}
\vskip2.1cm

The aim of this work is to consider the links between consistent and approximate descriptions of the
quantum-classical systems \cite{Kap}-\cite{GEP}, i.e. systems are composed of two interacting subsystems, one
of which behaves almost classically while the other requires a quantum
description. The formulation of a quantum-classical dynamics has attracted considerable interests
for the last decade \cite{Kap}, and as is well known, many conceptual difficulties arise in making this.
It will be shown that mixed quantum-classical dynamics is described by the
quantum-classical Heisenberg equation
for the evolution of observables or by the quantum-classical Liouville equation
for the evolution of states \cite{Kap},\cite{Proc},\cite{Ger1}-\cite{Ger3}.
These systems can be described by
the self-consistent Hamilton and Schr\"{o}dinger equations set only as an uncorrelated approximation.

A description of quantum-classical systems is formulated in
terms of two sets of objects: observables and states. The mean value (mathematical expectation) of observables
defines a duality between observables and states and as a
consequence there exist two approaches to the description of the
evolution.

An observable $A(X,\hat{X})$ of a quantum-classical system is a function of
canonical variables $X$ characterizing classical degrees of freedom and non-commuting
self-adjoint operators $\hat{X}$ satisfying canonical commutation relations \cite{Ger1},\cite{Ger3}.
For instance, the Hamiltonian of a quantum-classical system has the structure
\begin{eqnarray*}\label{H}
     H(X,\hat{X})= H_c(X)\hat{I}+ H_q(\hat{X})+ H_{\mathrm{int}}(X,\hat{X}),
\end{eqnarray*}
where $\hat{I}$ is a unit operator, $H_c, H_q, H_{\mathrm{int}}$ are correspondingly, the Hamiltonian
of the classical, quantum degrees of freedom and their interaction. Further, as an example, we will consider a
one-dimensional system consisting of two interacting classical and quantum
particles (with unit masses). In this case
\begin{eqnarray*}
H_c(X)=\frac{1}{2}\,p^2,\,\,\, H_q(\hat{X})=\frac{1}{2}\,\hat{p}^2, \,\,\, H_{\mathrm{int}}(X,\hat{X})=\Phi(q,\hat{q}),
\end{eqnarray*}
where $p,q$ are canonical variables of a classical particle, $\hat{q},\hat{p}$ are
self-adjoint operators satisfying canonical commutation relations and the function $\Phi$ is an interaction potential.

The mean value of the observable $A(X,\hat{X})$ at time instant $t\in\mathbb{R}$
is defined by the positive continuous linear functional on the space of observables
which can be defined in two different ways
\begin{eqnarray}\label{average}
\langle A \rangle(t)&&=\int dX \,\mathrm{Tr}\, A(t,X,\hat{X})D(0,X,\hat{X})=\\
&&=\int dX \,\mathrm{Tr}\, A(0,X,\hat{X})D(t,X,\hat{X}),\nonumber
\end{eqnarray}
where $A(0,X,\hat{X})$ is an observable at the initial instant $t=0$,
$D(0,X,\hat{X})$ is the density operator depending on classical canonical variables
and $\int dX \,\mathrm{Tr}\, D(0,X,\hat{X})=1$.

The evolution of observables $A(t)=A(t,X,\hat{X})$ is described by the initial-value problem
of the quantum-classical Heisenberg equation
\begin{eqnarray}\label{H-N1}
       &&\frac{\partial}{\partial t}A(t)=\mathcal{L}A(t),\\
       &&A(t)|_{t=0}=A(0),\nonumber
\end{eqnarray}
where the generator $\mathcal{L}$ of this evolution equation depends on the quantization rule of
a quantum subsystem. In the case of the Weyl quantization it has the form \cite{Ger1},\cite{Ger3}
\begin{eqnarray}\label{komyt}
 &&\mathcal{L}A(t)=-\frac{i}{\hbar}\big[A(t),H\big]+ \frac{1}{2}\big(\big\{A(t),H\big\}-\big\{H,A(t)\big\}\big),
\end{eqnarray}
where $H=H(X,\hat{X})$ is the Hamiltonian,  $[\,.\,,.\,]$ is a commutator of operators and $\{\,.\,,.\,\}$ is the Poisson brackets.
For some other quantization rules the operator $\mathcal{L}$ is defined in \cite{Ger1},\cite{Ger3}.

In the case of a one-dimensional system consisting of two interacting classical and quantum
particles in the configuration representation expression \eqref{komyt} has a form
\begin{eqnarray*}
      \big(\mathcal{L}A\big)(q,p;\xi,\xi')=&&-\frac{i}{\hbar}\big(-\frac{\hbar^2}{2}(-\Delta_{\xi}+\Delta_{\xi'})+
      \big(\Phi(\xi-q)-\Phi(\xi'-q)\big)\big)\,A(q,p;\xi,\xi') +\\
      &&\,+\,\big(\,p\frac{\partial}{\partial q}-\frac{\partial}{\partial q}\big(\Phi(\xi-q)-\Phi(\xi'-q)\big)
               \frac{\partial}{\partial p}\,\big)\,A(q,p;\xi,\xi'),
\end{eqnarray*}
where $A(q,p;\xi,\xi')$ is a kernel of the operator $A(X,\hat{X})$ in the configuration representation,
and in the Wigner representation it is as follow
\begin{eqnarray*}
  \big(\mathcal{L}A\big)(x_1,x_2)&&
 =\sum\limits_{j=1}^2 p_j\frac{\partial}{\partial q_j}A(x_1,x_2)-\frac{\partial}{\partial q_1}\Phi(q_1-q_2)
 \frac{\partial}{\partial p_1}A(x_1,x_2)+\\
 &&+\frac{i}{2\pi\hbar}\int d\eta d\xi \, e^{i(p_2-\xi)\eta}\,\big(\Phi \big(q_1-(q_2-\frac{\hbar}{2}\eta)\big)-\Phi \big(q_1-(q_2+\frac{\hbar}{2}\eta)\big)\big)\,A(x_1,q_2,\xi) ,
\end{eqnarray*}
where $x_i\equiv(q_i,p_i)\in\mathbb{R}\times\mathbb{R}$ and $A(x_1,x_2)$ is a symbol of the operator $A(X,\hat{X})$.

Usually the evolution of a quantum-classical system is
described, in the framework of evolution of states (the Schr\"{o}dinger picture of evolution),
by the initial-value problem dual to \eqref{H-N1}, namely, the quantum-classical Liouville equation
\cite{Kap},\cite{Ger1},\cite{Ger3}
\begin{eqnarray}\label{F-N1}
      &&\frac{\partial}{\partial t}D(t)=-\mathcal{L}D(t),\\
      &&D(t)|_{t=0}=D(0).\nonumber
\end{eqnarray}

First, let us construct the self-consistent field approximation of this equation.
For that we introduce the marginal states of classical and quantum subsystems correspondingly
\begin{eqnarray}
     \label{MC}  &&D(t,X) = \mathrm{Tr}\,D(t,X,\hat{X}),\\
     \label{MQ}  &&\hat{\rho}(t) = \int dX \,D(t,X,\hat{X})
\end{eqnarray}
and the correlation operator of classical and quantum subsystems
\begin{eqnarray}\label{C}
       &&g(t,X,\hat{X})= D(t,X,\hat{X})- D(t,X)\,\hat{\rho}(t).
\end{eqnarray}
If we assume that at any instant of time there are no correlations between classical and quantum particles,
i.e. it holds $g(t,X,\hat{X})=0$, then in such an uncorrelated approximation we derive from  \eqref{F-N1}
a self-consistent equations set of the Liouville and von Neumann equations for
marginal states \eqref{MC},\eqref{MQ} of classical and quantum  subsystems
\begin{eqnarray}\label{L-L}
      &&\frac{\partial}{\partial t}D(t,X)=\big\{H_c(X)+\mathrm{Tr}\,H_{\mathrm{int}}(X,\hat{X})\,\hat{\rho}(t)\,,\,D(t,X)\big\},\\
   \label{L-N} &&i\hbar\frac{\partial}{\partial t}\hat{\rho}(t)
       =\big[H_q(\hat{X})+\int dX H_{\mathrm{int}}(X,\hat{X})\,D(t,X)\,,\,\hat{\rho}(t)\big]
\end{eqnarray}
with initial data
\begin{eqnarray*}
       &&D(t)|_{t=0}=D(0), \\
       &&\hat{\rho}(t)|_{t=0}=\hat{\rho}(0).
\end{eqnarray*}

We note that the initial-value problem of equations \eqref{L-L},\eqref{L-N}
describes the evolution of all possible states of quantum-classical systems
if correlations between classical and quantum particles are neglected (the self-consistent field approximation).
This type of approximation can not be treated as the mean-field approximation
since we consider finitely many particles while the mean-field approximation assumes the transition to
the thermodynamic limit, i.e. it has a sense for infinitely many particles.
Rigorous results about the mean-field limit of dynamics of quantum many-particle systems in the framework of evolution of observables
and states are given in \cite{Ger4}.

Let at the initial instant the states of quantum and classical subsystems be pure states, i.e.
\begin{eqnarray} \label{in}
     &&D(0,X)=\delta\big(X-X_0\big),\\
     &&\hat{\rho}(0,\xi,\xi')=\Psi_0(\xi)\Psi^{*}_0(\xi'),\nonumber
\end{eqnarray}
where $\delta(X-X_0)$ is a Dirac measure, $X_0$ is the phase space point in which we measure
observables of classical subsystem, $\hat{\rho}(0,\xi,\xi')$ is a kernel (a density matrix) of
marginal density operator \eqref{MQ} in the configuration representation, which is
the projector operator on the vector $\Psi_0\in L^2$. We remark that the problem how to define a pure state
of the whole quantum-classical system is an open problem.

Then in the configuration representation the initial-value problem of equations \eqref{L-L},\eqref{L-N} is an
equivalent to the initial-value problem of a self-consistent set of the Hamilton and Schr\"{o}dinger equations
\begin{eqnarray}
  \label{H-S1}     &&\frac{\partial}{\partial t}X(t)
         =\big\{X(t)\,,\,H_c(X)+\int d\xi \,H_{\mathrm{int}}(X;\xi,\xi)\,|\Psi(t,\xi)|^2\big\},\\
  \label{H-S2}     &&i\hbar\frac{\partial}{\partial t}\Psi(t,\xi)=\int d\xi'\big(H_q(\xi,\xi')+
          H_{\mathrm{int}}(X(t);\xi,\xi')\big)\Psi(t,\xi')
\end{eqnarray}
with initial data
\begin{eqnarray*}
       && X(t)|_{t=0}=X,\\
       && \Psi(t,\xi)|_{t=0}=\Psi_0(\xi).
\end{eqnarray*}

For example, in the case of a one-dimensional system consisting of two interacting classical and quantum
particles (with unit masses) equations \eqref{H-S1},\eqref{H-S2} get the form
\begin{eqnarray*}\label{ex}
       &&\frac{d^2}{d t^2}Q(t)= -\, \frac{\partial}{\partial Q(t)}\int d\xi \, \Phi(Q(t)-\xi)\, |\Psi(t,\xi)|^2 ,\nonumber\\
       &&i\hbar \frac{\partial}{\partial t}\Psi(t,\xi)= - \frac{\hbar^2}{2}\frac{\partial^2}{\partial \xi^2}\Psi(t,\xi)+ \Phi(Q(t)-\xi)\Psi(t,\xi)\nonumber
\end{eqnarray*}
with initial data ($X\equiv(q,v)$)
\begin{eqnarray*}
       &&Q(t)|_{t=0}=q, \quad \frac{d}{d t}Q(t)|_{t=0}=v,\\
       &&\Psi(t,\xi)|_{t=0}=\Psi_0(\xi),
\end{eqnarray*}
where $\Phi(|Q(t)-\xi|)$ is a two-body interaction potential, $Q(t)$ is a position
at the instant $t$ of a classical particle in the space.
Such a self-consistent set of the Newton and Schr\"{o}dinger equations is usually used for the description
of the evolution of states of quantum-classical systems.

Now we consider the description of a quantum-classical system
in the uncorrelated approximation \eqref{C} between classical
and quantum particles in the framework of evolution of observables (the Heisenberg picture of evolution).

Assume that a quantum-classical system is in an uncorrelated pure state \eqref{in}, i.e.
\begin{eqnarray}\label{ups}
   &&  D(0,X;\hat{X})=\delta\big(X-X_0\big)P_{\Psi_0},
\end{eqnarray}
where $P_{\Psi_0}\equiv (\Psi_0,\,.\,)\Psi_0$  (or in Dirac notation: $P_{\Psi_0}\equiv |\Psi_0\rangle\langle\Psi_0|$) is a one-dimensional projector
onto a unit vector ${\Psi_0}$ from a Hilbert space.
In the configuration representation a kernel of operator \eqref{ups} has the form
\begin{eqnarray*}
&&D(0,X;\xi,\xi')=\delta\big(X-X_0\big)\Psi_0(\xi)\Psi^{*}_0(\xi').
\end{eqnarray*}
Then according to \eqref{average} in this approximation the evolution of the canonical observables
of quantum and classical particles is described by the Heisenberg equations set
\begin{eqnarray}\label{L-Ho}
   \label{L-Ho1}    &&\frac{\partial}{\partial t}X(t)\hat{I}=
          \big\{X(t)\hat{I}\,,\, H_c(X)\hat{I}+ H_{\mathrm{int}}(X,\hat{X})\big\},\\
   \label{L-Ho2}    &&i\hbar\frac{\partial}{\partial t}\hat{X}(t)=
          \big[\hat{X}(t)\,,\, H_q(\hat{X})+ H_{\mathrm{int}}(X,\hat{X})\big]
\end{eqnarray}
with initial data
\begin{eqnarray*}
&& X(t)|_{t=0}=X,\\&& \hat{X}(t)|_{t=0}=\hat{X}.
\end{eqnarray*}

For example, in the case of a one-dimensional system consisting of two interacting classical and quantum
particles equations \eqref{L-Ho1},\eqref{L-Ho2} can be rewritten for pairs of canonically conjugated variables
$X(t)=\big(Q(t),P(t)\big)$ and $\hat{X}(t)=\big(\hat{Q}(t),\hat{P}(t)\big)$ in the following form
\begin{eqnarray*}\label{QC-h1}
       &&\frac{d}{d t}Q(t)=P(t),\\
       &&\frac{d}{d t}P(t)\hat{I}= - \frac{\partial}{\partial Q(t)}\Phi(Q(t),\hat{Q}(t)),\\
       &&\frac{d}{d t}\hat{Q}(t)=\hat{P}(t),\\
       &&\frac{d}{d t}\hat{P}(t)= - \Phi'(Q(t),\hat{Q}(t)),
\end{eqnarray*}
where the function $\Phi'$ is the derivative of the function $\Phi$.
These equations can also be rewritten similar to
\eqref{H-S1},\eqref{H-S2} as equations in the configuration representation.
In the Wigner representation they take the following fascinating form
\begin{eqnarray*}\label{QC-h1}
       &&\frac{d}{d t}Q_1(t)=P_1(t),\\
       &&\frac{d}{d t}P_1(t)= - \frac{\partial}{\partial Q_1(t)}\Phi(Q_1(t)-Q_2(t)),\\
       &&\frac{d}{d t}Q_2(t)=P_2(t),\\
       &&\frac{d}{d t}P_2(t)= - \frac{\partial}{\partial Q_2(t)}\Phi(Q_1(t)-Q_2(t))
\end{eqnarray*}
with initial data
\begin{eqnarray*}
&& X_1(t)|_{t=0}=x_1,\,\,\,
 X_2(t)|_{t=0}=x_2,
\end{eqnarray*}
where $X_i(t)\equiv\big(Q_i(t),P_i(t)\big),\, i=1,2$, and $X_2(t)\equiv\big(Q_2(t),P_2(t)\big)$  are symbols of the operators $\hat{X}(t)=\big(\hat{Q}(t),\hat{P}(t)\big)$.
We note that in this representation a symbol of uncorrelated pure state \eqref{ups} defines by the following Wigner function
\begin{eqnarray*}
&&D(0,x_1,x_2)=(2\pi)^{-1} \delta\big(x_1-x_0\big)\int d\xi\, \Psi_0\big(q_2+\frac{1}{2}\hbar\,\xi\big)
\Psi^{*}_0\big(q_2-\frac{1}{2}\hbar\,\xi\big)e^{-i\xi p_2}
\end{eqnarray*}
and mean value \eqref{average} of an observable
is given by the following functional
\begin{eqnarray*}
\langle A \rangle(t)=(2\pi\hbar)^{-1}\int dx_1 dx_2\, A(t,x_1,x_2)\,D(0,x_1,x_2).
\end{eqnarray*}

In summary, mixed quantum-classical dynamics is described by the
quantum-classical Heisenberg equation \eqref{H-N1}
in the framework of the evolution of observables or by the quantum-classical Liouville equation \eqref{F-N1}
in the framework of the evolution of states. A quantum-classical system  can be described by
the self-consistent Hamilton \eqref{H-S1} and Schr\"{o}dinger \eqref{H-S2} equations set
or by the self-consistent Hamilton \eqref{L-Ho1} and Heisenberg \eqref{L-Ho2} equations set and state \eqref{ups} only
in an uncorrelated approximation \cite{Ger2},\cite{Ger3}.

\end{document}